\documentclass[aps,pra,twocolumn,showpacs,prabib]{revtex4}
\usepackage{graphicx}
\usepackage{amsmath}
\usepackage{amssymb,wasysym}
\usepackage{amsfonts}
\usepackage{bm,verbatim}

\newcommand{\bea}{\begin{eqnarray}}
\newcommand{\eea}{\end{eqnarray}}
\newcommand{\be}{\begin{equation}}
\newcommand{\ee}{\end{equation}}

\begin{document}
\title{Elastic scattering of a quantum matter-wave bright soliton on a barrier}

\author{Christoph Weiss}
\affiliation{Institut f\"ur Physik, 
Universit\"at Oldenburg, 26111 Oldenburg, Germany}
\affiliation{Department of Physics, Durham University, Durham DH1 3LE, United Kingdom}

\author{Yvan Castin}
\affiliation{Laboratoire Kastler Brossel, \'Ecole Normale
Sup\'erieure, UPMC and CNRS, 24 rue Lhomond, 75231 Paris Cedex 05, France}

\begin{abstract}
We consider a one-dimensional matter-wave bright soliton, corresponding to the ground bound state
of $N$ particles of mass $m$ having a binary attractive delta potential interaction on the open line.
For a full $N$-body quantum treatment, we derive several results 
for the scattering of this quantum soliton on a short-range, bounded from below, external potential, 
restricting to the low energy, elastic regime where the centre-of-mass kinetic energy of the
incoming soliton is lower than the internal energy gap of the soliton, that is the minimal energy
required to extract particles from the soliton.
\end{abstract}

\date{\today}

\maketitle

\section{Introduction}

The ultracold atomic Bose gases have proved to be very flexible physical systems, 
where both the dimensionality and the interaction strength can be adjusted at will.
By trapping Bose-condensed atoms in an optical waveguide that freezes their transverse motion in its ground state,
one obtains ultracold one-dimensional Bose gases. By further making the effective one-dimensional atomic interaction 
attractive, one can produce matter-wave bright solitons, which are bound states of matter 
with typically thousands of particles \cite{Salomon,Hulet,Cornish}.
This opens up a new field with exciting possibilities, the field of coherent 
matter-wave optics with massive objects.
Even a soliton of light atoms such as ${}^7$Li is typically more massive than the big 
organic molecules (such as fullerenes)
used in interferometric experiments \cite{Arndt}, and it has the 
advantage of having much larger
centre-of-mass de Broglie wavelength, since the atomic gases can be prepared 
in the nK temperature range \cite{Ketterle}.

The disadvantage (or depending on the perspective, the additional feature) of the matter-wave soliton is that
it is a quite fragile object: The ground state soliton is separated by a continuum of fragmented solitons
by a small energy gap $\Delta$, with $\Delta/k_{\rm B}$ typically sub-microKelvin.
For the scattering of a quantum soliton on a barrier to be guaranteed to be elastic 
by energy conservation, one
has to restrict the kinetic energy of the centre of mass of the soliton to 
values below the gap $\Delta$. This elastic scattering regime 
is quite intriguing and was recently
considered in proposals for 
production of real space Schr\"odinger-cat-like  states by 
coherent splitting by a laser barrier
of the centre-of-mass wavepacket into transmitted and reflected components 
\cite{PRL2009,Cederbaum}
and for Anderson localisation of quantum solitons in a disordered potential
\cite{Cord2009} \footnote{The production of entangled states by scattering of two quantum solitons
was studied in \cite{Lewenstein}.}.
On the experimental side, scattering of a soliton on a barrier 
is under experimental investigation, for the moment out of the elastic scattering regime, 
with fragmentation of the soliton into two main pieces \cite{Hulet_private}.

Here we restrict to the elastic scattering regime on a localized potential barrier:
inside or close to the potential, 
the system can virtually access internal excited states (where the soliton
is fragmented) but it fully occupies the ground state soliton at asymptotically
large distances from the barrier, so that scattering of the soliton with incoming
centre-of-mass wavevector $K$ is characterized by the transmission amplitude
$t$ and the reflection amplitude $r$ with $|r|^2+|t|^2=1$. As an initial wavepacket may be expanded
over such stationary scattering states, its time-dependent wavefunction
away from the barrier can be deduced from the $K$-dependent $t$ and $r$ amplitudes.

Whereas the classical field (or Gross-Pitaevskii) equation was extensively
used to study soliton dynamics and fragmentation in external potentials
\cite{solitonGP}, it does not look appropriate in the elastic scattering regime.
First, the Gross-Pitaevskii equation does not provide a full quantum-mechanical
treatment of the centre-of-mass motion. In the absence of an external potential,
it predicts the existence of localized stationary solutions, whereas the 
centre-of-mass position necessarily spreads ballistically in time in the quantum
world \cite{LesHouches}. In the scattering by a barrier, it cannot
describe Schr\"odinger-cat-like states, where the unfragmented soliton
has some non-zero probability amplitude to be to the left (resp. to the right)
of the barrier \cite{PRL2009}.
Secondly, the classical soliton misses the rigidity of the quantum soliton
at the heart of elastic scattering: in the classical field theory,
the moving soliton can in principle always slow down 
by radiating at infinity an arbitrary small
amount of energy, without violating energy conservation,
whereas in the quantum theory, the number of particles radiated
to infinity (that carry away an energy at least $\Delta$) is quantized.

We thus have to use the quantum field theory, which constitutes a full
many-body problem when the number of bosons $N$ is large.
In the absence of a barrier, it was solved with the Bethe ansatz
generalized to complex quasi-momenta, both for the ground state
\cite{Guire} and for the excited states \cite{CRAS,Caux1,Caux2};
the many-particle ground state in the presence of a harmonic trap 
was investigated in \cite{DH2012}.
In the presence of a barrier, the Bethe ansatz is not applicable, the exact
$N$-body solution is not known and one has to resort to approximations. 
When the barrier is broad as compared to the soliton size, 
it is natural to introduce the average $\bar{V}(X)$ of the external potential
experienced by the $N$ bosons over the density profile $\rho(x|X)$ 
of the ground state soliton with centre of mass localized in $X$. 
Then one writes a Schr\"odinger equation for a centre-of-mass wavefunction
$\Phi(X)$, treated as a single particle of mass $M=Nm$ ($m$ is the mass
of a single boson) moving in the potential $\bar{V}(X)$.
This intuitive approximation was used e.g.\ in \cite{PRL2009,Cord2009}.

The scope of the present paper is to provide tools to construct
this approximation, to control it with rigorous error bounds on the transmission
and reflection amplitudes, and to go one step beyond it in the large-$N$ limit.
In Sec.~\ref{sec:effec_pot}, we define the problem; using a projector technique,
we show that the centre-of-mass wavefunction $\Phi(X)$ can be given a precise
meaning and that it obeys, in the elastic regime, an exact Schr\"odinger-like
equation with an effective potential that, in addition to $\bar{V}(X)$,
contains a non-local and energy dependent contribution $\delta \mathcal{V}$
originating from all possible virtual fragmentations of the soliton.
In Sec.\ref{sec:brack} we derive a simple upper bound on the matrix elements
of $\delta \mathcal{V}$, which allows to derive upper bounds (already used
in \cite{PRL2009}) on
the error on $t$ and $r$ due to the omission of $\delta \mathcal{V}$;
in the case of a very narrow potential barrier, such as a repulsive Dirac
delta, we show how to improve the procedure to get usable upper bounds.
In Sec.\ref{sec:lnl} we determine from Bogoliubov theory the leading order
contribution to $\delta\mathcal{V}$ in the large $N$ limit, with this limit
constructed in such a way that $\bar{V}(X)$ remains fixed.
In Sec.~\ref{sec:BO} we again consider the large-$N$ limit case with a 
Born-Oppenheimer-like approach, the heavy particle being the centre of mass, 
and we identify
a regime where it approximately coincides with the Bogoliubov result
of Sec.\ref{sec:lnl}.
In Sec.\ref{sec:appli} we give simple applications of the formalism.
We conclude in Sec.~\ref{sec:conclusion}.

\section{Definition of the problem and the effective potential for elastic scattering}
\label{sec:effec_pot}

\subsection{Hamiltonian and free space properties}

We consider $N$ spinless bosons of mass $m$ moving quantum-mechanically
on the open one-dimensional line.
The bosons have an attractive Dirac pair interaction
characterized by the negative coupling constant $g$ \cite{Olshanii}, and each boson
is subjected to a localized potential $U(x)$, that is $U(x)$ rapidly tends to
zero for $|x|\to +\infty$.  
The $N$-body Hamiltonian $H$ is the sum of the free space
Hamiltonian $H_0$ and of the external potential Hamiltonian $V$.
In first quantized form:
\bea
H &=& H_0 + V \\
H_0 &=& \sum_{i=1}^{N} \frac{p_i^2}{2m} +\sum_{1\leq i< j\leq N} 
g\delta(x_i-x_j) \\
V &=& \sum_{i=1}^{N} U(x_i),
\label{eq:defV}
\eea
where $x_i$ is the spatial coordinate of the $i^{\rm th}$ boson and
$p_i$ is its momentum operator.

The free space Hamiltonian $H_0$ can be diagonalised with the Bethe ansatz
\cite{CRAS,Caux1,Caux2}. Another key feature of $H_0$, 
that we shall use extensively, is the separability of the centre-of-mass 
degrees of freedom (associated to the centre-of-mass position $X$) 
from the internal degrees of freedom (whose $N-1$ spatial coordinates 
can be expressed in terms of the $(x_i)_{1\leq i\leq N}$ through Jacobi
formulas \cite{IFT}, that are not required here).
This gives a tensorial product structure to the Hilbert space between 
the centre-of-mass variable and the internal variables, and it corresponds
to the following splitting between the centre-of-mass kinetic energy operator
and the internal Hamiltonian $H_{\rm int}$:
\be
H_0 = \frac{P^2}{2M} + H_{\rm int}
\label{eq:sepH0}
\ee
with $M=Nm$
is the total mass and $P=\sum_i p_i$ is the total momentum operator.
The internal Hamiltonian $H_{\rm int}$ does not depend at all 
on the centre-of-mass variable
\footnote{This perfect decoupling would not take place in a quantization box with periodic
boundary conditions, as the boundary conditions for the internal variables
would then depend on the centre-of-mass momentum \cite{Lieb}.}.
It has only one discrete eigenstate, its ground state $|\phi\rangle$ of
eigenenergy $E_0(N)$ given by \cite{Guire}
\be
E_0(N) = -\frac{mg^2}{24\hbar^2} (N-1)N(N+1).
\label{eq:E0N}
\ee
Considered as a function of the $x_i$'s, $\phi$ is also the ground state of
$H_0$ since it corresponds to a centre of mass at rest,
$ H_0 | \phi\rangle = E_0(N) |\phi\rangle$.
It has a simple expression in terms of the single particle coordinates \cite{Guire}
\be
\phi(x_1,\ldots,x_N) = \mathcal{N} e^{-\frac{m|g|}{2\hbar^2} 
\sum_{1\leq i < j \leq N} |x_i-x_j|
}
\label{eq:valphi}
\ee
with a normalisation condition also easily expressed, by fixing the 
centre-of-mass position to the origin of coordinates:
\be
\int dx_1\ldots dx_N\, \delta\left(\sum_{i=1}^{N}x_i/N\right)|\phi(x_1,\ldots,x_N)|^2 = 1,
\ee
which leads to \cite{Calogero,CRAS}
\be
|\mathcal{N}|^2 \left(\frac{\hbar^2}{m|g|}\right)^{N-1}\frac{N}{(N-1)!} =1.
\ee
Apart from this discrete eigenstate, 
the spectrum of $H_{\rm int}$ is a continuum separated from
$E_0(N)$ by a gap $\Delta$, corresponding to any possible fragmentation of the ground state
soliton into smaller solitons (including single particles) with arbitrary 
centre-of-mass momenta.
From the full spectrum obtained by the Bethe ansatz \cite{CRAS,Caux1,Caux2}, and
using $E_0(n_1)+E_0(n_2)> E_0(n_1+n_2)$ for $n_1,n_2>0$,  one finds 
\be
\label{eq:DeltaN}
\Delta= E_0(N-1)+E_0(1)-E_0(N)=
\frac{m g^2}{8\hbar^2} N(N-1),
\ee
i.e.\  $\Delta$ is the energy required to extract a particle of vanishing relative
momentum from the $N$-particle soliton.

In the presence of the external localized potential $U(x)$, the centre of mass of the gas
experiences scattering and is no longer decoupled. Let us assume first that $U(x)$ is everywhere
non-negative, so that no boson can remain trapped in the potential.
Then at low enough energy $E$ such that
\be
E_0(N) < E < E_0(N)+ \Delta
\ee
the eigenstate $|\psi\rangle$ of $H$ of energy $E$,
\be
0= (E-H) |\psi\rangle,
\label{eq:Seq}
\ee
has a simple structure far away from the external potential,
corresponding to elastic scattering of the ground state soliton:
Because of energy conservation, at a centre-of-mass position $X\to \pm\infty$,
the internal state is in its ground state $\phi$ and the centre-of-mass
wavefunction assumes the usual asymptotic form of a single particle scattering state.
Introducing the positive $K$ such that
\be
E = E_0(N) + \frac{\hbar^2 K^2}{2M} 
\ \ \mbox{with}\ \ \frac{\hbar^2 K^2}{2M} < \Delta,
\ee
we thus have the boundary conditions (see Fig.~\ref{fig:schema}):
\bea
\psi(x_1,\ldots,x_N) &\sim& \Phi(X)\, \phi(x_1,\ldots,x_N) \\
\label{eq:alinfinip}
\Phi(X) &\underset{X\to-\infty}{\sim} & e^{iKX}+re^{-iKX} \\
\Phi(X) &\underset{X \to+\infty}{\sim} & t e^{iKX},
\label{eq:alinfinim}
\eea
where $|r|^2+|t|^2=1$. 
As we shall see, a meaning can be given to the centre-of-mass
wavefunction $\Phi(X)$ of the soliton at all $X$, not simply
at infinity.  The goal of the present work 
is to calculate approximately the reflection amplitude $r$ and
the transmission amplitude $t$, and to control the resulting error.

The previous physical reasoning has to be adapted when $U(x)$ presents weakly negative parts,
that may support bound states, in which case the scattering state could be fragmented,  e.g.\
it could correspond to a $(N-n)$-particle soliton flying away, with $n$ bosons trapped within the external potential ($0<n<N$).
A $n$-particle bound state, being an eigenstate of the free space Hamiltonian plus external potential Hamiltonian,
has an energy $E_{\rm bound}$ necessarily larger than the sum of the minimal eigenvalues of each Hamiltonian,
$E_{\rm bound}\geq E_0(n) + n \inf_x U(x)$. The energy of the fragmented scattering state 
is thus larger than $E_0(N-n)+ E_0(n) + n \inf_x U(x) > E_0(N) + \Delta + N  \inf_x U(x)$.
Fragmented scattering states are thus forbidden by energy conservation over the energy range
\be
E_0(N) < E < E_0(N) +  \Delta + N  \inf_x U(x),
\label{eq:range}
\ee
a constraint over $E$ and $\inf_x U(x)$ that we assume to be satisfied (to guarantee purely elastic soliton scattering)
and that will be recovered by a purely mathematical reasoning.

\subsection{The effective potential}

An exact rewriting of Schr\"odinger's equation within a restricted subspace is
obtained with the action of projectors on the resolvent 
$G(z)=1/(z-H)$ of the Hamiltonian \cite{CCT}, leading to an effective Hamiltonian.
Using the tensorial product structure of the Hilbert space between the 
centre-of-mass and the internal variables,
we define the operator projecting orthogonally the internal variables 
onto their ground state $|\phi\rangle$:
\be
\mathcal{P} = 1_{\rm CM} \otimes |\phi\rangle \langle\phi|
\ee
where $1_{\rm CM}$ stands for the operator identity over the 
centre-of-mass variables. The supplementary orthogonal projector is
\be
\mathcal{Q} = 1 -\mathcal{P}.
\ee
Then for any complex and non-real number $z$, we obtain the exact expression 
\cite{help}
\be
\mathcal{P} G(z) \mathcal{P} = \frac{\mathcal{P}}{z\mathcal{P} -\mathcal{P} H\mathcal{P}
-\mathcal{P}V\mathcal{Q}\frac{\mathcal{Q}}{z\mathcal{Q}-\mathcal{Q}H\mathcal{Q}}
\mathcal{Q}V\mathcal{P}}.
\label{eq:exexp}
\ee
To access the scattering state of energy $E$, one should usually take the limit
$z=E+i\epsilon$, $\epsilon\to 0^+$, because the operator
$z\mathcal{Q}-\mathcal{Q}H\mathcal{Q}$ is usually not invertible for $z=E$ (within
the subspace over which $\mathcal{Q}$ projects).
But we shall now restrict to a situation where this operator is invertible because
it is strictly negative.
To this end, we assume that the external potential $U(x)$ is bounded from
below,
\be
\inf_x U(x) >-\infty
\ee
Then the spectrum (abbreviated as Spec) of the Hermitian operator
$\mathcal{Q}H\mathcal{Q}$ (within the subspace over which
$\mathcal{Q}$ projects) is also bounded from below,
\be
\inf \mbox{Spec}\, \mathcal{Q}H\mathcal{Q} \geq N\inf_x U(x) + E_0(N) + \Delta.
\label{eq:inf_qhq}
\ee
To ensure that $\mathcal{Q}H\mathcal{Q} -E \mathcal{Q}$ is strictly positive
(within the subspace over which $\mathcal{Q}$ projects),
we thus impose
\be
\frac{\hbar^2 K^2}{2M} < \Delta + N\inf_x U(x),
\label{eq:constraint_E}
\ee
which reproduces the physical result (\ref{eq:range}).
The action of the projector $\mathcal{P}$ onto the eigenstate $|\psi\rangle$
gives
\be
\mathcal{P} |\psi\rangle = |\Phi\rangle\otimes |\phi\rangle.
\label{eq:action}
\ee
The wavefunction 
\be
\Phi(X)=\langle X|\Phi\rangle = (\langle X|\otimes \langle \phi|) | \psi\rangle
\ee
plays a crucial role, it is the centre-of-mass wavefunction within the subspace 
where the internal variables
are in their ground state (that is in the minimal energy $N$-particle soliton).
In other words, $\Phi(X)$ is the centre-of-mass wavefunction of the soliton.
In what follows, we shall use a shorthand notation of the type
$|\Phi\rangle = \langle \phi | \psi\rangle$, where the tensorial product
structure between centre-of-mass and internal variables is implicitly 
assumed.  In terms of the original variables $x_i$, it is expressed as
\begin{multline}
\Phi(X)= \int dx_1\ldots dx_N \,\delta(X-\sum_{i=1}^{N} x_i/N)  \\
\phi^*(x_1,\ldots,x_N) \psi(x_1,\ldots,x_N).
\end{multline}
For $z\to E$, the effective Hamiltonian appearing in the denominator of Eq.~(\ref{eq:exexp})
is Hermitian under condition (\ref{eq:constraint_E}), and we find the exact 
Schr\"odinger-like equation for $|\Phi\rangle$:
\be
\frac{\hbar^2 K^2}{2M} |\Phi\rangle =
\left[\frac{P^2}{2M} + \bar{V}(\hat{X}) + \delta\mathcal{V}\right] |\Phi\rangle,
\label{eq:heff}
\ee
where $P=-i\hbar \partial_X$ and, as we shall discuss, $\bar{V}$ is given by (\ref{eq:barV}) and $\delta\mathcal{V}$ is given
by (\ref{eq:defdV}).
This result can also be obtained by a direct calculation without introducing the resolvent:
One applies the projector $\mathcal{P}$ and the projector $\mathcal{Q}$ to
Schr\"odinger's equation (\ref{eq:Seq}), and one inserts the closure relation $\mathcal{P}+\mathcal{Q}=1$
to the right of $(E-H)$ to obtain
\begin{eqnarray}
E \mathcal{P} |\psi\rangle &=& \mathcal{P} H \mathcal{P} |\psi\rangle + (\mathcal{P} H \mathcal{Q}) 
\mathcal{Q} |\psi\rangle \\
(E\mathcal{Q}-\mathcal{Q} H \mathcal{Q}) \mathcal{Q} |\psi\rangle  &=& \mathcal{Q} H \mathcal{P} |\psi\rangle
\end{eqnarray}
Under the condition (\ref{eq:constraint_E}) we multiply the second equation
by the inverse of the operator $E\mathcal{Q}-\mathcal{Q} H \mathcal{Q}$ (within the subspace
over which $\mathcal{Q}$ projects) and we report the resulting value of $\mathcal{Q} |\psi\rangle$
within the first equation. After some rewriting using in particular (\ref{eq:action})
we recover (\ref{eq:heff}).

The first contribution to the effective potential in (\ref{eq:heff})
is very intuitive, and is a simple
function of the centre-of-mass position $X$,
\be
\bar{V}(X) = \langle \phi| V | \phi\rangle = \langle V\rangle_X
\label{eq:barV}
\ee
where we have introduced $\langle \ldots \rangle_X$ 
the expectation value in the internal ground
state for a fixed value $X$ of the centre-of-mass position.
For a general observable $O$ that is diagonal in terms of the original $x_i$ variables, one has:
\begin{multline}
\langle O\rangle_X =
\int dx_1\ldots dx_N \, \delta(X-\sum_{i=1}^{N} x_i/N) \\
\times O(x_1,\ldots, x_N) |\phi(x_1,\ldots,x_N)|^2. 
\label{eq:defmX}
\end{multline}
We shall see that $\bar{V}(X)$ is then simply
the convolution of the external potential $U(x)$ 
with the density profile $\rho(x|0)$
of the ground state  soliton whose centroid is fixed at the origin of coordinates.

The second contribution to the effective
potential in Eq.~(\ref{eq:heff}) is both non-local and dependent on the
scattering state energy $E$:
\be
\delta \mathcal{V} = \langle \phi| V \mathcal{Q} \frac{\mathcal{Q}}{E\mathcal{Q}-\mathcal{Q}H\mathcal{Q}} 
\mathcal{Q} V |\phi\rangle.
\label{eq:defdV}
\ee
Its evaluation cannot even by performed with the Bethe ansatz, due to the presence of
the external potential $V$ in the denominator. A first strategy is to simply neglect $\delta \mathcal{V}$
as compared to $\bar{V}$ in (\ref{eq:heff}), as already done in \cite{PRL2009,Cord2009}, which intuitively 
should be accurate in the large $N$ limit and/or when $U(x)$ is broad as compared to the soliton size $\xi$.
In Sec.~\ref{sec:brack}, rigorous bounds on the resulting error on the soliton transmission and reflection
amplitudes are given. A second strategy is to calculate the leading large-$N$ asymptotic expression
of $\delta \mathcal{V}$ within Bogoliubov theory, as done in Sec.~\ref{sec:lnl}, or to rely on a simpler
Born-Oppenheimer-like approximation, as done in Sec.~\ref{sec:BO},
that can be subsequently used in a numerical solution of (\ref{eq:heff}).

\subsection{How to calculate $\bar{V}(X)$}

To obtain an operational expression for $\bar{V}(X)$,
we introduce the mean density of particles in the ground state soliton
for a fixed position of the centre of mass,
\be
\rho(x|X) = \langle \hat{\rho}(x)\rangle_X,
\ee
where the operator giving the density in point $x$ is
\be
\hat{\rho}(x) = \sum_{i=1}^{N} \delta(x_i-x).
\ee
Using Eq.~(\ref{eq:defmX}) we thus reach the intuitive result
\be
\bar{V}(X) = \int_\mathbb{R} dx\, U(x) \rho(x|X)= \int_\mathbb{R} 
dx\, U(X-x) \rho(x|0)
\label{eq:Vbarop}
\ee
using the translational invariance $\rho(x|X)=\rho(x-X|0)$ and the fact that
$\rho(x|0)$ is an even function of $x$. 
This also enables an exact evaluation of $\bar{V}(X)$,
since $\rho(x|X)$ was calculated with
the Bethe ansatz in \cite{Calogero,CRAS}:
\be
\rho(x|X) = \frac{N!^2}{N\xi} \sum_{k=0}^{N-2} \frac{(-1)^k(k+1)}{(N-2-k)!(N+k)!}
e^{-(k+1)|x-X|/\xi},
\label{eq:calogero}
\ee
where $\xi$ is the spatial width of the classical field (Gross-Pitaevskii) soliton,
\be
\xi = \frac{\hbar^2}{m |g| N}.
\label{eq:xi}
\ee
A large-$N$ expansion can be obtained from (\ref{eq:calogero}) \cite{Calogero,
Castin_EPJB}: For $N\to\infty$ with $Ng$ (and thus $\xi$) fixed,
\be
\rho(x|0) = N \phi_0^2(x) -\xi^2 \frac{d^2}{dx^2} [\phi_0^2(x)] + o(1),
\ee
where the classical field soliton single particle wavefunction
(normalized to unity) is given by
\be
\phi_0(x) = \frac{1}{2\xi^{1/2}} \frac{1}{\cosh(x/2\xi)}.
\label{eq:def_phi0}
\ee
For $N\to +\infty$ with fixed $\xi$, a double integration by part leads to
\begin{multline}
\bar{V}(X) =  
\int_{\mathbb{R}} dx\, N \phi_0^2(x)  \left[U(x+X) - 
\frac{\xi^2}{N} U''(x+X)\right]  \\
+ o(U).
\label{eq:devVbar}
\end{multline}

\section{Bracketing the transmission and reflection amplitudes}
\label{sec:brack}

In the simplest treatment of elastic soliton scattering, one simply neglects the contribution 
$\delta \mathcal{V}$ in (\ref{eq:heff}) \cite{PRL2009,Cord2009}. 
The challenge is to be able to put a bound on the corresponding error
performed on the transmission and reflection coefficients. To this end, we derive
an upper bound on the modulus of the matrix elements of $\delta \mathcal{V}$.
Then we derive upper bounds on the contribution of $\delta \mathcal{V}$ to the
transmission and reflection coefficients, first in a minimal version (where the bounds can be 
directly evaluated from existing Bethe ansatz results), and second in a refined version applicable
also to arbitrarily narrow external potentials (such as a repulsive delta potential).

\subsection{Upper bound on the matrix elements of $\delta \mathcal{V}$}
\label{subsec:upper_bound}

Let us consider two kets $|\Phi_1\rangle$ and $|\Phi_2\rangle$ 
for the centre-of-mass degrees of freedom. The corresponding wavefunctions need
not be square integrable but the following kets should be normalisable,
\bea
\label{eq:defu1}
|u_1\rangle &=& \mathcal{Q} V |\Phi_1\rangle \otimes |\phi\rangle  \\
|u_2\rangle &=& \mathcal{Q} V |\Phi_2\rangle \otimes |\phi\rangle .
\label{eq:defu2}
\eea
The matrix element of $\delta \mathcal{V}$ may thus be written as
\be
\langle \Phi_1|\delta \mathcal{V}|\Phi_2\rangle = \langle u_1| (- \mathcal{G}) | u_2\rangle
\ee
where we have introduced the Hermitian operator
\be
\mathcal{G} = \frac{\mathcal{Q}}{\mathcal{Q}H\mathcal{Q}-E\mathcal{Q}}.
\label{eq:defG}
\ee
According to relations (\ref{eq:inf_qhq},\ref{eq:constraint_E}), 
the operator $\mathcal{G}$ is positive
(within the subspace over which $\mathcal{Q}$ projects).
From the Cauchy-Schwarz inequality,
\be
|\langle u_1|\mathcal{G}|u_2\rangle|^2 \leq \langle u_1|u_1\rangle
\langle u_2| \mathcal{G}^2 |u_2\rangle.
\ee
From Eq.~(\ref{eq:inf_qhq}) we find, e.g.\ injecting a closure relation in
the eigenbasis of $\mathcal{Q}H\mathcal{Q}$,
\be
\langle u_2| \mathcal{G}^2 |u_2\rangle \leq \frac{\langle u_2|u_2\rangle}
{[N\inf_x U(x)+\Delta-\hbar^2 K^2/2M]^2}.
\ee
Calculating the norm squared of the $|u_i\rangle$, we find
\be
\langle u_i|u_i\rangle = \langle \Phi_i| w(\hat{X}) | \Phi_i\rangle
\ee
where we have introduced the positive quantity
\bea
w(X) &=&  \langle V^2\rangle_X - \langle V\rangle_X^2 
\label{eq:defW}
\eea
with $\langle\ldots\rangle_X$ defined in (\ref{eq:defmX}) and $V$ given by (\ref{eq:defV}).

In conclusion, the non-local contribution $\delta \mathcal{V}$ to the effective Hamiltonian
appearing in (\ref{eq:heff}) can be bounded by the local positive potential (that we call
{\sl error potential})
\be
W(X) = \frac{w(X)}{N\inf_x U(x) + \Delta -\hbar^2 K^2/2M}
\label{eq:defgW}
\ee
in the following sense:
\be
|\langle \Phi_1| \delta \mathcal{V}|\Phi_2\rangle|  \leq 
\left[\langle \Phi_1|W(\hat{X})|\Phi_1\rangle
\langle \Phi_2|W(\hat{X})|\Phi_2\rangle\right]^{1/2}.
\label{eq:maj}
\ee

\subsection{Upper bound on the error on the scattering coefficients}
\label{subsec:theoreme_faible}

In the simplest approximation, one neglects $\delta \mathcal{V}$ in Eq.~(\ref{eq:heff}) and one calculates
the wavefunction $\Phi_0(X)$ corresponding to scattering of the soliton
centre of mass onto the potential $\bar{V}(X)$ with incoming wavevector $K>0$ (i.e.\ for a soliton coming from the left):
\be
\frac{\hbar^2 K^2}{2M} \Phi_0(X) = -\frac{\hbar^2}{2M} \Phi_0''(X) 
+\bar{V}(X)\Phi_0(X).
\ee
It leads to a transmission amplitude $t_0$
and a reflection amplitude $r_0$, see Fig.~\ref{fig:schema}.

In the exact treatment, keeping $\delta \mathcal{V}$, the transmission and reflection amplitudes
are $t$ and $r$, see Fig.~\ref{fig:schema}. We introduce the two positive quantities:
\bea
\epsilon &=& \frac{M \langle \Phi_0|W(\hat{X})|\Phi_0\rangle}{\hbar^2 K |t_0|}
\label{eq:epsilon} \\
\label{eq:eta}
\eta &=& \frac{\langle \tilde{\Phi}_0|W(\hat{X})|\tilde{\Phi}_0\rangle}
{\langle \Phi_0|W(\hat{X})|\Phi_0\rangle}
\eea
where the scattering solution onto $\bar{V}(X)$ for an incoming wave with negative wavevector $-K$ (i.e.\ for a
soliton coming from
the right) can be expressed in terms of $\Phi_0(X)$:
\be
\tilde{\Phi}_0(X)= \frac{\Phi_0^*(X)-r_0^*\Phi_0(X)}{t_0^*}.
\label{eq:phi0t}
\ee
Then we have the rigorous result:

\begin{figure}[t]
\centerline{\includegraphics[width=0.9\linewidth,clip=]{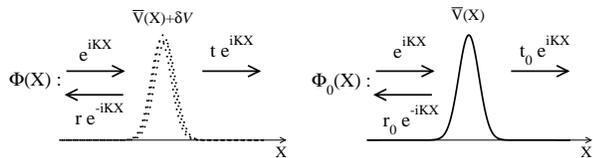}}
\caption{Left panel: In the regime of elastic scattering of a $N$-boson bright quantum soliton on a localized external potential,
the centre-of-mass wavefunction $\Phi(X)$ obeys boundary conditions for $X\to\pm\infty$
with transmission and reflection amplitudes $t$ and $r$ satisfying $|t|^2+|r|^2=1$. 
The effective potential seen by $\Phi$
involves a non-local part $\delta \mathcal{V}$ corresponding to virtual soliton fragmentation,
see Eq.~(\ref{eq:heff}). Right panel: In the simplest approximation, one neglects the non-local part
$\delta \mathcal{V}$, only keeping the average $\bar{V}(X)$ of the single particle external potential $U(x)$
over the density profile $\rho(x|X)$ of the soliton with centre-of-mass in $X$. The resulting approximate transmission
and reflection amplitudes are $t_0$ and $r_0$. Bounds on $|t-t_0|$ and $|r-r_0|$ are derived in
Sec.~\ref{sec:brack}, and tools to evaluate the leading order corrections due to $\delta \mathcal{V}$ 
in the large $N$ limit are developed in Sec.~\ref{sec:lnl} and in Sec.~\ref{sec:BO}.}
\label{fig:schema}
\end{figure}

\noindent {\sl Theorem: If $\epsilon \eta^{1/2} < 1/2$ then}
\bea
\label{eq:bound_tg}
||t|-|t_0|| \leq |t-t_0| &\leq & \frac{|t_0|\epsilon\eta^{1/2}}{1-2\epsilon\eta^{1/2}}  \\
||r|-|r_0|| \leq |r-r_0| &\leq & \frac{|t_0|\epsilon}{1-2\epsilon\eta^{1/2}}.
\label{eq:bound_rg}
\eea
The first inequalities result from the triangular inequality.
The proof of the second inequalities is given in the Appendix \ref{appen:theorem}.
In the case of an even external potential, one has simplified relations,
since $\eta=1$:

\noindent {\sl Theorem: If $U(x)$ is even and $\epsilon<1/2$, then}
\bea
\label{eq:bound_t}
||t|-|t_0|| \leq  |t-t_0| &\leq & \frac{|t_0|\epsilon}{1-2\epsilon}  \\
||r|-|r_0|| \leq |r-r_0| &\leq & \frac{|t_0|\epsilon}{1-2\epsilon}.
\label{eq:bound_r}
\eea

\subsection{General results for the error potential $W(X)$}

We now review some exact results derived in \cite{Castin_EPJB}
on the function $w(X)$ appearing in the numerator of the error potential $W(X)$,
see Eqs.~(\ref{eq:defW},\ref{eq:defgW}).
The function $w(X)$ can be expressed in terms 
of the static structure factor of the ground state soliton for a fixed position $X$ of its centre of mass:
\bea
S(x,y|X) &=& \langle \hat{\rho}(x) \hat{\rho}(y)\rangle_X \\
&=& \delta(x-y) \rho(x|X) + \rho(x,y|X)
\eea
where $\rho(x,y|X)$ is the pair distribution function of the soliton of 
centre of mass localized in $X$.
One has indeed
\begin{multline}
w(X) = \int_{\mathbf{R}^2}  dxdy\, U(x+X) U(y+X)  \\
\times [S(x,y|0)-\rho(x|0)\rho(y|0)].
\label{eq:WvsS}
\end{multline}
This writing immediately reveals that $w(X)$ depends on correlations that would not be accurately
treated in the classical field (Gross-Pitaevskii) approximation.
From the Bethe ansatz wavefunction (\ref{eq:valphi}) of the soliton,
Fourier transforms of $\rho(x|0)$ and $\rho(x,y|0)$ were expressed as sums of $N$ and $O(N^2)$ terms
in \cite{Castin_EPJB}, which allows an exact calculation of $w(X)$.

Useful limiting cases were also studied in \cite{Castin_EPJB}. 
For a {\sl broad} external potential, with $U(x)$ varying slowly over the width of the soliton,
that can be estimated by the width $\xi$ of the classical soliton given in Eq.~(\ref{eq:xi}),
$w(X)$ is essentially proportional to the square of the second order derivative of $U$,
\be
w(X) \simeq C(N)\, \xi^4 [U''(X)]^2.
\label{eq:wbblne}
\ee
The coefficient $C(N)$ only depends on $N$, it is given as a sum of $O(N^3)$ terms,
and it has the large-$N$ asymptotic behaviour
\be
C(N) \underset{N\to \infty}{\sim}  N \left[\frac{2\pi^2}{3} + 4 \zeta(3)\right]
\label{eq:CNg}
\ee
where $\zeta$ is the Riemann Zeta function.
For a {\sl narrow} external potential, centered at the origin of coordinates with a width
much smaller than the soliton width $\xi$, one has for any $N$:
\be
\label{eq:w_si_pot_etroit}
w(X) \simeq \rho(X|0) \int_{-\infty}^{+\infty} dx\, U^2(x).
\ee
Finally, irrespective of the width of $U(x)$, one can simplify Eq.~(\ref{eq:WvsS})
in the large $N$ limit by using an asymptotic expression for the pair distribution
function \cite{Castin_EPJB}:
\begin{multline}
w(X) \sim 2 N \xi^4 \int_{\mathbb{R}} dx \int_x^{+\infty} \!\! dy\, U''(X+x\xi) \,U''(X+y\xi) \\
\times \frac{2+y-x}{(e^y+1)(e^{-x}+1)}
\end{multline}
where $Ng$ (and thus $\xi$) are kept fixed while $N\to +\infty$.

\subsection{An improved bracketing applicable to a Dirac external potential}

A limitation of the transmission and reflection coefficient bracketing 
of subsection \ref{subsec:theoreme_faible} is that it becomes useless when
the external potential $U(x)$ is too narrow.
For example, in the limiting case of a repulsive Dirac potential, $U(x) = v \delta(x)$, $v>0$,
it is apparent that the quantity $\langle V^2\rangle_X$ in Eq.~(\ref{eq:defW})
in infinite, since it contains a sum over all the particles of $\delta^2(x_i)$.
As a consequence, the quantity $\epsilon$ defined in Eq.~(\ref{eq:epsilon})
is $+\infty$ and the theorem applicability condition $\epsilon < 1/2$ is not satisfied.

Here, we show that a slight improvement of the derivation allows to remove
this limitation. The resulting bracketing is thus more stringent, the price to pay
being that the new upper bound on $|t-t_0|$ is more difficult to evaluate
in practice.

One simply uses the fact that
\be
\mathcal{Q} H \mathcal{Q} - E \mathcal{Q} \geq \mathcal{Q} H_{\rm int} \mathcal{Q} + N \inf_x U(x) \mathcal{Q}
- E \mathcal{Q} 
\label{eq:boundQHQ}
\ee
where as usual, for two Hermitian operators $A$ and $B$, $A \geq B$ means that
the operator $A-B$ is non-negative, that is $\langle u| (A-B) |u\rangle \geq 0$ for any ket $|u\rangle$.
Eq.~(\ref{eq:boundQHQ}) results from the fact that the centre-of-mass
kinetic energy operator $P^2/2M$ is non-negative, and that each operator $U(x_i)$ 
is larger than or equal to $\inf_x U(x)$. Since we still impose Eq.~(\ref{eq:constraint_E}) on the energy $E$,
and since $\mathcal{Q} H_{\rm int} \mathcal{Q} \geq [E_0(N)+\Delta] \mathcal{Q}$,
the operator in the right-hand side of Eq.~(\ref{eq:boundQHQ}) is positive (within the subspace over which $\mathcal{Q}$ projects).

For two positive Hermitian operators $A$ and $B$ such that $A \geq B$, one has that
\footnote{If $A\geq B >0$, then $B^{-1/2} A B^{-1/2} > 1$, where $1$ is the identity. If a self-adjoint operator
$C$ satisfies $C \geq 1$, then $C^{-1} \leq 1$, as may be checked in the eigenbasis of $C$. Then
$B^{1/2} A^{-1} B^{1/2} \leq 1$ which implies Eq.~(\ref{eq:relinv}).}
\be
\label{eq:relinv}
B^{-1}  \geq A^{-1}.
\ee
Applying this relation for $A$ and $B$ being the left-hand side and the right-hand side operators
in Eq.~(\ref{eq:boundQHQ}), considered within the subspace over which  $\mathcal{Q}$ projects, 
one finds that
\be
0\leq \mathcal{G} \leq \frac{\mathcal{Q}}{N \inf_x U(x)\mathcal{Q} +  \mathcal{Q} H_{\rm int} \mathcal{Q} - E \mathcal{Q}}
\label{eq:majG}
\ee
where $\mathcal{G}$ is defined in Eq.~(\ref{eq:defG}). From the Cauchy-Schwarz inequality, writing $\mathcal{G} = \left(\mathcal{G}^{1/2}\right)^2$,
we have for arbitrary kets $|u_{1,2}\rangle$ such that $\mathcal{G}^{1/2} |u_{1,2}\rangle$ are 
normalisable:
\be
|\langle u_1 | \mathcal{G} |u_2\rangle|^2 \leq \langle u_1| \mathcal{G} |u_1\rangle \langle u_2|\mathcal{G} |u_2\rangle.
\ee
We apply this inequality to the kets $|u_1\rangle$ and $|u_2\rangle$ defined in Eqs.(\ref{eq:defu1},\ref{eq:defu2}), that do not need to be 
normalisable, and we use the upper bound on the operator $\mathcal{G}$ to obtain
for arbitrary centre-of-mass wavefunctions $\Phi_{1,2}(X)$ (not diverging too fast at infinity):
\be
|\langle \Phi_1| \delta \mathcal{V}| \Phi_2\rangle| \leq \left[\langle \Phi_1| W_{\rm imp}(\hat{X}) |\Phi_1\rangle
\langle \Phi_2| W_{\rm imp}(\hat{X}) |\Phi_2\rangle \right]^{1/2},
\label{eq:betterboundingdV}
\ee
where the improved error potential is positive:
\be
\label{eq:Wimp}
W_{\rm imp}(X) = \langle \phi| V \mathcal{Q} \frac{\mathcal{Q}}{N \inf_x U(x)\mathcal{Q} +  \mathcal{Q} H_{\rm int} \mathcal{Q} - E \mathcal{Q}}
\mathcal{Q} V |\phi\rangle.
\ee
Thanks to the occurrence of the internal kinetic energy operator of the particles within $H_{\rm int}$ in the denominator,
this error potential remains finite even when the barrier is a repulsive $\delta$ potential.

The reasoning of subsection \ref{subsec:upper_bound}  may then be reproduced, replacing the error potential $W$ by the improved
one. Similarly to Eqs.(\ref{eq:epsilon},\ref{eq:eta}) we thus define
\bea
\epsilon_{\rm imp} &=& \frac{M \langle \Phi_0|W_{\rm imp}(\hat{X})|\Phi_0\rangle}{\hbar^2 K |t_0|}
\label{eq:epsilon_imp} \\
\label{eq:eta_imp}
\eta_{\rm imp} &=& \frac{\langle \tilde{\Phi}_0|W_{\rm imp}(\hat{X})|\tilde{\Phi}_0\rangle}
{\langle \Phi_0|W_{\rm imp}(\hat{X})|\Phi_0\rangle}
\eea
where, as in subsection \ref{subsec:upper_bound}, $\Phi_0(X)$ and $\tilde{\Phi}_0(X)$
are the centre-of-mass scattering wavefunctions with incoming wavevector $K$ and $-K$ respectively,
for the potential $\bar{V}(X)$.
One then has:\\
{\sl Improved theorem: If $\epsilon_{\rm imp} \eta_{\rm imp}^{1/2} < 1/2$ then}
\bea
\label{eq:boundimp_t}
||t|-|t_0|| \leq |t-t_0| &\leq & \frac{|t_0|\epsilon_{\rm imp}\eta_{\rm imp}^{1/2}}{1-2\epsilon_{\rm imp}\eta_{\rm imp}^{1/2}}  \\
||r|-|r_0|| \leq |r-r_0| &\leq & \frac{|t_0|\epsilon_{\rm imp}}{1-2\epsilon_{\rm imp}\eta_{\rm imp}^{1/2}},
\label{eq:boundimp_r}
\eea
where $r,t$ are the exact reflection and transmission coefficients, and $r_0,t_0$ are
the reflection and transmission coefficients for $\Phi_0$, that is for the potential $\bar{V}(X)$.
As in subsection \ref{subsec:upper_bound}, a simpler form is obtained for an even external potential $U(x)=U(-x)$,
in which case $\eta_{\rm imp}=1$.

\subsection{General results for the improved error potential $W_{\rm imp}(X)$}

A general calculation of $W_{\rm imp}(X)$ with the Bethe ansatz, amounting
to evaluating an internal dynamic structure factor of the ground state 
soliton with fixed centre-of-mass position, may be doable with the techniques
developed in \cite{Caux1,Caux2} but this is beyond the scope
of this paper. 
On the contrary, a large $N$ limit (for fixed $Ng$ and $\xi$) is straightforward
to obtain from the Bogoliubov technique exposed in Sec.~\ref{sec:lnl}:
In Eq.~(\ref{eq:dVBog}) one simply has to omit the centre-of-mass kinetic energy
term and to replace $\bar{V}(X)$ by the lower bound $N \inf_x U(x)$:
\be
W_{\rm imp}(X) \underset{N\to +\infty}{\sim} 
\int_\mathbb{R} \frac{dk}{2\pi} \frac{|\Gamma_k(X)|^2}{N \inf_x U(x) 
+\frac{\hbar^2 k^2}{2m} + \Delta - \frac{\hbar^2 K^2}{2M}}
\ee
where the amplitude $\Gamma_k(X)$ is given by (\ref{eq:Gamma})
and $\Delta\simeq \hbar^2/(8 m\xi^2)$, see (\ref{eq:DeltaN}).

When the external potential is a repulsive delta potential,
$U(x)=v \delta(x)$, $v>0$, with $\inf_x U(x)=0$, the integral over $k$
can be calculated:
\be
W_{\rm imp}(X) \sim \frac{2 N m v^2 \xi}{\gamma \hbar^2} \phi_0^2(X)
\left[1-\frac{32(\gamma+2)}{(\gamma+1)^2} \xi^3 \phi_0'^{2}(X)\right]
\label{eq:Wimpdelta}
\ee
where $\gamma \in (0,1)$ is such that
\be
\gamma^2 = 1- \frac{\hbar^2 K^2}{2M\Delta}.
\ee
As expected, (\ref{eq:Wimpdelta}) diverges when the incoming
centre-of-mass kinetic energy tends to the gap $\Delta$, but it diverges
as $1/\gamma$, whereas the error potential $W(X)$ generically diverges
as $1/\gamma^2$ for a non-negative $U(x)$, see Eq.~(\ref{eq:defgW}).

This indicates that the improved bound can have some interest
also for a broad barrier. In the large $N$ limit, when $U(X)$ has a width
$b\gg \xi$, we find using (\ref{eq:Gam_large}) and assuming for simplicity
that $\inf_x U(x)=0$:
\begin{multline}
W_{\rm imp}(X) \underset{b\gg\xi}{\simeq} \frac{8\pi^2N \xi^4}{\Delta}
[U''(X)]^2  \\
\times \int_{\mathbb{R}} \frac{dk}{2\pi} [(1+k^2)^2(\gamma^2+k^2)\cosh^2(\pi k/2)]^{-1},
\end{multline}
where the integral may be expressed analytically if necessary, in particular in terms of the derivative
of the digamma function, using the residue theorem.

\section{Large $N$ limit of the effective potential for a $O(1/N)$ barrier}
\label{sec:lnl}

We calculate a large $N$ expansion of the non-local part $\delta \mathcal{V}$
of the effective potential in Eq.~(\ref{eq:heff}),
using Bogoliubov theory,
in the case where the external potential $U$ experienced by each particle
scales as $1/N$ and the soliton width $\xi$ is fixed (because $N g$ is
fixed). This physically convenient scaling with $N$ 
ensures that the potential $\bar{V}$ has a well-defined non-zero
limit for $N\to +\infty$.

\subsection{Bogoliubov theory in brief}

We use Bogoliubov theory to dress with quantum fluctuations the classical
soliton of single particle wavefunction (\ref{eq:def_phi0}).
Since $\phi_0$ is centered at the origin of coordinates, we shall shift the positions
of the particles as  $x_i\to x_i + X$ where $X$ is the fixed position of the 
centre of mass of the quantum soliton.
In the number conserving theory \cite{Gardiner,CastinDum}, one splits the bosonic field operator as
\be
\hat{\psi}(x)= \hat{a}_0 \phi_0(x) + \hat{\psi}_\perp(x)
\ee
where $\hat{a}_0$ annihilates a particle in the mode $\phi_0$ and the field $\hat{\psi}_\perp(x)$
is orthogonal to the field $\phi_0(x)$. We introduce the  modulus-phase representation 
\cite{Girardeau,Nieto}, which is an excellent
approximation  for large $N$ (when the probability of having an empty  mode $\phi_0$ is negligible):
\be
\hat{a}_0 = e^{i\hat{\theta}} \hat{n}_0^{1/2},
\ee
where the Hermitian phase operator $\hat{\theta}$ is conjugate to the number operator $\hat{n}_0=\hat{a}_0^\dagger \hat{a}_0$,
$[\hat{n}_0,\hat{\theta}]=i$.  The phase $\hat{\theta}$ is formally eliminated by its inclusion with the field
$\hat{\psi}_\perp$ in the number conserving field
\be
\hat{\Lambda}(x) = e^{-i\hat{\theta}} \hat{\psi}_\perp(x).
\ee
Conservation of the total number of particles allows to eliminate $\hat{n}_0$ in terms of $\hat{\Lambda}$
and of the total number operator $\hat{N}$. In the large $N$ limit (with $N g$ fixed), it is found
that $\hat{\Lambda}=O(1/\xi^{1/2})$ whereas $\hat{a}_0$ scales as $N^{1/2}$, which allows a systematic expansion of the Hamiltonian
in powers of $\hat{\Lambda}$. Keeping terms up to order 
$O(\hat{\Lambda}^2)$ in $H_0$  leads to the
Bogoliubov approximation for $N$ particles,
\be
H_0 \simeq E_0^{\rm Bog}(N) + \frac{\hat{P}^2}{2N m}  + \int_{\mathbf{R}} \frac{dk}{2\pi} \epsilon_k
\hat{b}_k^\dagger \hat{b}_k
\ee
with the gapped Bogoliubov spectrum in terms of the quasiparticle wavevector $k$ and the Gross-Pitaevskii
chemical potential $\mu_0$:
\be
\epsilon_k = |\mu_0| + \frac{\hbar^2k^2}{2m}  \ \ \ \mbox{\rm where} \ \ \ \mu_0 = -\frac{\hbar^2}{8m\xi^2}.
\ee
The quasi-particle annihilation and creation operators $\hat{b}_k$ and
$\hat{b}_k^\dagger$ obey the usual bosonic commutation relations on the
open line, $[\hat{b}_k,\hat{b}_{k'}^\dagger] = 2\pi \delta (k-k')$.
Due to the translational symmetry breaking, a Goldstone mode appears, with a massive term $\propto \hat{P}^2$
in the Hamiltonian, the field variable $\hat{P}$ (scaling as $N^{1/2}$)
representing at the Bogoliubov level the total momentum
of the system, and being conjugate to the field variable $\hat{Q}$ (scaling as $1/N^{1/2}$) 
giving at the Bogoliubov level the fluctuations of the centre-of-mass position of the system: This
reproduces the structure of Eq.~(\ref{eq:sepH0}).  
The modal field expansion is then
\bea
\hat{\Lambda}(x) &=& -N^{1/2} \phi_0'(x) \hat{Q} + \frac{i}{\hbar N^{1/2}} x\phi_0(x) \hat{P} \nonumber \\
&+& \int_{\mathbf{R}} \frac{dk}{2\pi} [u_k(x) \hat{b}_k + v_k^*(x) \hat{b}_k^\dagger].
\label{eq:modal}
\eea
The Bogoliubov mode functions are known exactly \cite{Kaup} and are given with the present notations
in \cite{Castin_EPJB}. They are orthogonal to $\phi_0$, and one has also that $u_k+v_k$ is orthogonal
to $x \phi_0(x)$.
This is apparent on the useful form:
\be
\phi_0(x) [u_k(x)+v_k(x)] = \frac{4\xi^2}{(1+2i|k|\xi)^2}
\frac{d^2}{dx^2} \left[e^{ikx} \phi_0(x)\right].
\label{eq:upv_utile}
\ee

\subsection{Bogoliubov expression of $\delta \mathcal{V}$}

To calculate $\delta \mathcal{V}$ given by Eq.~(\ref{eq:defdV}), 
we first have to express the operator $V$ in the Bogoliubov framework.
To the same level of approximation as for the Hamiltonian $H_0$, that is neglecting terms that are cubic
or more in $\hat{\Lambda}$, we obtain 
\begin{multline}
\sum_{i=1}^{N}  U(X+x_i)  \simeq (N-\int_{\mathbb{R}} dx\, \hat{\Lambda}^\dagger\hat{\Lambda}) 
\int_{\mathbb{R}} dx\, \phi_0^2(x)U(X+x)  \\
+N^{1/2} \int_{\mathbb{R}} dx\, [\hat{\Lambda}(x)+\hat{\Lambda}^\dagger(x)] \phi_0(x) U(X+x) \\
+\int_{\mathbb{R}} dx\, U(X+x) \hat{\Lambda}^\dagger(x) \hat{\Lambda}(x).
\label{eq:VBog}
\end{multline}
It is clear that, in the Bogoliubov framework,  applying the projector $\mathcal{P}$ of the full quantum theory
amounts to projecting onto the vacuum $|0\rangle_{\rm Bog}$
of all the quasiparticle
annihilation operators $\hat{b}_k$, so that applying $\mathcal{Q}$ amounts to projecting onto the subspace with at least
one quasiparticle excitation. In the Bogoliubov evaluation of $\mathcal{Q} V |\phi\rangle$, 
the leading term explicitly scaling as $N$ in Eq.~(\ref{eq:VBog}) gives a vanishing contribution, so we keep
the subleading $N^{1/2}$ term of (\ref{eq:VBog}). Similarly, due to the projector $\mathcal{Q}$,
we keep only the contributions involving $\hat{b}_k^\dagger$ in Eq.~(\ref{eq:modal}) to obtain
\be
\mathcal{Q} V |\phi\rangle \simeq \int_{\mathbb{R}} \frac{dk}{2\pi} \, \Gamma_k(X) \hat{b}_k^\dagger |0\rangle_{\rm Bog}
\ee
with the amplitudes depending parametrically on $X$:
\begin{multline}
\label{eq:Gamma}
\Gamma_k(X) = 
N^{1/2} \int_{\mathbb{R}} dx\, [u_k^*(x)+v_k^*(x)] \phi_0(x) U(X+x) \\
= \frac{4\xi^2 N^{1/2}}{(1-2i|k|\xi)^2} \int_{\mathbb{R}} dx\, 
e^{-ikx} \phi_0(x) U''(X+x)
\end{multline}
where we used (\ref{eq:upv_utile}).
The integral over $x$ is typically cut to $|x|\lesssim \xi$ by the rapidly decreasing function
$\phi_0(x)$.
For an external potential narrower than $\xi$, $|\Gamma_k(X)|$ varies with $X$ at the length scale $\xi$.
For a broad external potential, varying at a scale $b\gg \xi$, we expect that 
$\Gamma_k(X)$ varies with $X$ at the length scale $b$. This can be made quantitative
by expanding $U''(X+x)$ in the integrand of Eq.~(\ref{eq:Gamma}) to 
zeroth order in $x$: 
\be
\Gamma_k(X) \underset{b\gg\xi} {\simeq} 
\frac{4\pi \xi^{5/2}N^{1/2} U''(X)}{(1-2i|k|\xi)^2\cosh(\pi k\xi)}.
\label{eq:Gam_large}
\ee

It remains to estimate the denominator $E\mathcal{Q} -\mathcal{Q} H \mathcal{Q}$ with Bogoliubov theory.
Within the subspace with one Bogoliubov excitation, the leading term explicitly
scaling as $N$ in Eq.~(\ref{eq:VBog}) gives a non-zero contribution
which is actually $O(1)$ since $U=O(1/N)$. This contribution does not affect the quasiparticle, it is a scalar depending
on $X$ only, and it simply corresponds to the mean-field (Gross-Pitaevskii) approximation
for $\bar{V}(X)$. 
The subleading term in Eq.~(\ref{eq:VBog}), which changes the number of quasiparticles by
$\pm 1$, is more involved since it couples the single-quasiparticle subspace
to the two-quasiparticle subspace. However,
its contribution is $O(N^{1/2}U)=O(1/N^{1/2})$ and may be neglected at the present order.

We finally obtain in the large $N$ limit, for $U=O(1/N)$ and fixed $Ng$, the leading term of $\delta \mathcal{V}$, scaling
as $1/N$:
\begin{multline}
\delta \mathcal{V} \sim \int_{\mathbb{R}} \frac{dk}{2\pi} \Gamma_k^*(X) \\
\times \left\{
E-\left[\frac{P^2}{2M} + E_0(N) + \epsilon_k + \bar{V}(X)\right]\right\}^{-1} \Gamma_k(X).
\label{eq:dVBog}
\end{multline}
Here $P=-i\hbar \partial_X$ is the centre-of-mass momentum operator 
of the full quantum theory.
The expansion of $\bar{V}$ up to the same order (neglecting $o(1/N)$) is 
directly given by (\ref{eq:devVbar}).
One may then solve numerically the resulting approximate form
of Eq.~(\ref{eq:heff}), which is made delicate by the non-local
nature of (\ref{eq:dVBog}). More simply, one may  treat (\ref{eq:dVBog})
as a first order perturbation in the $\bar{V}$ scattering problem using
the formulation of the Appendix  \ref{appen:theorem} [see below Eq.~(\ref{eq:sol_for})], to obtain:
\bea
\label{eq:dtpo}
t-t_0 &\simeq & -\frac{i M}{\hbar^2 K} \langle \tilde{\Phi}_0^* | \delta \mathcal{V} |\Phi_0\rangle \\
r-r_0 &\simeq & -\frac{i M}{\hbar^2 K} \langle \Phi_0^* | \delta \mathcal{V} |\Phi_0\rangle,
\label{eq:drpo}
\eea
the ket $|\Phi_0^*\rangle$ corresponding to the wavefunction $\Phi_0^*(X)$.

\section{Born-Oppenheimer-like approach} 
\label{sec:BO}

In molecular physics, one often uses the so-called Born-Oppenheimer approximation:
One diagonalises the electronic problem for fixed positions of the nuclei,
obtaining a ground state electronic energy that then serves as a potential for 
the nuclei \cite{BO,IFT}.
It is natural to try to apply a similar approach to our soliton scattering problem.
The ``heavy" particle then corresponds to the centre of mass of the soliton,
and the ``light" particles are the internal degrees of freedom of the soliton.
We thus split the $N$-body Hamiltonian as $H=H^{\rm heavy} + H^{\rm light}$,
with
\bea
H^{\rm heavy} &=& \frac{P^2}{2M} + N U(X) \\
H^{\rm light} &=& H_{\rm int} + \sum_{i=1}^{N} [U(x_i)-U(X)].
\eea
In $H^{\rm heavy}$, we have included the external potential that the soliton 
would feel if all the particles were localized in the centre-of-mass position $X$.
We expect this term $N U(X)$ to constitute already a good approximation 
when the external potential is much broader than the soliton width $\xi$.
To go beyond this zeroth order approximation, the idea is to calculate the ground
state energy $E_0^{\rm light}(X)$ of $H^{\rm light}$ for a fixed value of the 
centre-of-mass position.  This energy then provides a correction to the potential 
$N U(X)$ experienced by the centre of mass of the soliton.
An important condition for the Born-Oppenheimer approximation to hold
is that $E_0^{\rm light}(X)$ is well separated from the excited state energies
of $H^{\rm light}$ for a fixed $X$, so that the presence of the solitonic internal gap 
again plays an important role here.

More precisely, we call $|\chi(X)\rangle$ the ground state of the internal Hamiltonian
$H^{\rm light}$ corresponding to the eigenvalue $E_0^{\rm light}(X)$, and we put
forward the Born-Oppenheimer-like ansatz for the $N$-body state vector:
\be
\langle X|\psi\rangle_{\rm BO} = \Phi(X)  |\chi(X)\rangle
\label{eq:BOa}
\ee
where the ket $|X\rangle$ represents 
the centre of mass perfectly localized in $X$ and the internal
ket $|\chi(X)\rangle$, normalized to unity, parametrically depends on $X$
\footnote{In terms of the Jacobi internal coordinates 
$y_1,\ldots, y_{N-1}$, Eq.~(\ref{eq:BOa}) means that
$\psi_{\rm BO}(x_1,\ldots,x_N) = \Phi(X) \chi(y_1,\ldots,y_{N-1};X)$
with $X=\frac{1}{N} \sum_{i=1}^{N} x_i$.}.
We then insert the ansatz into Schr\"odinger's equation $E |\psi\rangle = H |\psi\rangle$
and project with $\langle \chi(X)|$ to obtain
\footnote{The natural choice that $|\chi(X)\rangle$ has a real wavefunction leads
to $\langle \chi(X)| \frac{d}{dX} |\chi(X)\rangle = 0$ since $\langle\chi(X)|\chi(X)\rangle=1$
for all $X$.}
\begin{multline}
E \Phi(X) = -\frac{\hbar^2}{2M} \frac{d^2}{dX^2} \Phi(X) 
+ 
\Big[N U(X) + E_0^{\rm light}(X) \\
-\frac{\hbar^2}{2M} \langle \chi(X)| \frac{d^2}{dX^2} |\chi(X)\rangle\Big] \Phi(X).
\end{multline}
Note that we keep here the so-called Born-Oppenheimer diagonal correction 
coming from the $X$ dependence of the internal state in the ansatz. 

In practice, to evaluate $E_0^{\rm light}(X)$ and the corresponding  eigenvector,
we use perturbation theory, treating $\sum_{i=1}^{N} [U(x_i)-U(X)]$
as a perturbation of $H_{\rm int}$. For example, to second order in this perturbation,
we obtain:
\begin{multline}
E_0^{\rm light}(X) \simeq E_0(N)-N U(X) + \langle \phi|V|\phi\rangle  \\
+ \langle \phi| V \mathcal{Q} \frac{\mathcal{Q}}{E_0(N) \mathcal{Q} - \mathcal{Q} H_{\rm int} \mathcal{Q}}
\mathcal{Q} V | \phi\rangle,
\label{eq:pertur}
\end{multline}
where, as in the previous sections, the internal ket $|\phi\rangle$ is the free space ground state
soliton of energy $E_0(N)$, $\mathcal{P}$ projects orthogonally onto $|\phi\rangle$ and the supplementary
projector $\mathcal{Q}=1-\mathcal{P}$ projects onto the internal excited states
of the system. Remarkably, the third term in the right-hand side of Eq.~(\ref{eq:pertur})
exactly coincides with $\bar{V}(X)$.
To first order in the perturbation theory, the internal ground state in presence of the external
potential is
\be
|\chi(X)\rangle \simeq \mathcal{F}(X) \left[|\phi\rangle 
+ \frac{\mathcal{Q}}{E_0(N)\mathcal{Q}-\mathcal{Q} H_{\rm int} \mathcal{Q}} \mathcal{Q} V |\phi\rangle\right].
\ee
The normalization factor $\mathcal{F}$ should for consistency only weakly deviate from
unity, which imposes a limit on the strength of the external potential.

We apply the above approximation scheme in the large $N$ limit, where it makes the most sense,
fixing $Ng$ (and thus the soliton width $\xi$).
We can then use the Bogoliubov approach, and
following the lines of section \ref{sec:lnl}:
\begin{multline}
E_0^{\rm light}(X) \simeq E_0^{\rm Bog}(N) -NU(X) + \bar{V}(X)  \\ + \int_{\mathbb{R}} \frac{dk}{2\pi} 
\frac{|\Gamma_k(X)|^2}{-\epsilon_k},
\label{eq:dVBO}
\end{multline}
where $\Gamma_k(X)$ is given by Eq.~(\ref{eq:Gamma}).
Also, the Born-Oppenheimer diagonal correction is approximated with Bogoliubov theory
as
\begin{multline}
-\frac{\hbar^2}{2M} \langle \chi (X)| \frac{d^2}{dX^2} |\chi(X)\rangle 
\simeq \frac{\hbar^2}{2M} \int_\mathbb{R} \frac{dk}{2\pi} \frac{|\frac{d}{dX}\Gamma_k(X)|^2}{\epsilon_k^2}.
\end{multline}
It is about $N$ times smaller than the Bogoliubov term appearing in $E_0^{\rm light}(X)$, see the last term in Eq.~(\ref{eq:dVBO}),
and we neglect it in the large $N$ limit.
To summarize, in the Born-Oppenheimer approximation for large $N$, we find for the soliton wavefunction 
an equation of the form (\ref{eq:heff}) with the non-local $\delta\mathcal{V}$ approximated
by the local form [after use of Eq.~(\ref{eq:Gamma})]
\begin{widetext}
\begin{multline}
\delta\mathcal{V}_{\rm BO} \simeq \int_{\mathbb{R}} \frac{dk}{2\pi}
\frac{|\Gamma_k(X)|^2}{-\epsilon_k} =
-\frac{N\xi}{8\Delta} \int_{\mathbb{R}^2}dxdy\, \phi_0(x)\phi_0(y) U''(X+x) U''(X+y) e^{-|x-y|/2\xi}
[(x-y)^2+6\xi |x-y| +12 \xi^2]
\label{eq:BOdV}
\end{multline}
\end{widetext}
The integral can be evaluated for a delta external potential, $U(x)=v\delta(x)$, leading to
\footnote{This can be transformed using $4\xi^2\phi_0''(X)=\phi_0(X)-8\xi\phi_0^3(X)$.}:
\be
\label{eq:vbopud}
\delta\mathcal{V}_{\rm BO} \simeq -\frac{5}{4} N \frac{m v^2}{\hbar^2} \left\{
\xi \phi_0^2(X) + \frac{48}{5} \xi^5 \left[\phi_0''(X)\right]^2\right\}.
\ee

It is interesting to compare Eq.~(\ref{eq:BOdV}) to the result Eq.~(\ref{eq:dVBog}) that was obtained in
a different way. At first sight, Eq.~(\ref{eq:dVBog}) and Eq.~(\ref{eq:BOdV})
look widely different, because of the more complicated energy denominator in
Eq.~(\ref{eq:dVBog}) that involves both $P^2/2M$ and $\bar{V}(X)$, rather then
a simple c-number quantity such as $\epsilon_k$.
We have identified a limiting case where the two expressions are close,
when the typical wavevector $K$ of $\Phi(X)$ is much larger than $1/\xi$, and $\delta \mathcal{V}$ 
has a small perturbative effect on the scattering state $\Phi$.
Since the relevant $k$ are $O(1/\xi)$ in the integral over $k$,
$\Gamma_k(X)$ varies at a length scale of order $\xi$ or larger,
see discussion below Eq.~(\ref{eq:Gamma}), 
whereas $\Phi(X)$ varies over a much smaller length scale $1/K$.
The spatial derivatives of $\Gamma_k(X) \Phi(X)$ are thus well approximated by taking
the derivatives of $\Phi(X)$ only, e.g.
\be
P \Gamma_k(X) |\Phi\rangle \underset{K\xi \gg 1}{\simeq}  \Gamma_k(X) P |\Phi\rangle,
\ee
where $P=-i\hbar \partial_X$ is the centre-of-mass momentum operator.
We can then approximately commute the energy denominator with $\Gamma_k(X)$
in Eq.~(\ref{eq:dVBog}). The last step is to realise that, if $\delta\mathcal{V}$ is small enough,
$\Phi(X)$ will be close to the scattering state of energy $E$ for the
potential $E_0(N)+\bar{V}(X)$ so that
\be
\left\{
E-\left[\frac{P^2}{2M} + E_0(N) + \epsilon_k + \bar{V}(X)\right]\right\}^{-1} 
|\Phi\rangle \simeq  -\frac{1}{\epsilon_k} |\Phi\rangle
\ee
and we recover the Born-Oppenheimer result Eq.~(\ref{eq:BOdV}).

\section{Applications of the formalism}
\label{sec:appli}

\subsection{Centre-of-mass wavepacket splitting}

We apply our formalism to the proposal of \cite{PRL2009}
for the production of Schr\"odinger's cat-like states by elastic scattering of a soliton 
on a barrier: One sends the ground state soliton with a quasi-monochromatic
centre-of-mass wavepacket, that is centered in $K$-space around $\bar{K}$ with a  width $\Delta K \ll \bar{K}$,
on a barrier of adjusted height such that the 
transmission and reflection amplitudes have the same modulus $1/\sqrt{2}$. This prepares the
gas in a coherent superposition of all the particles being to the right and to the left
of the barrier with equal probability amplitudes. 
For the experimental decoherence rate estimated in \cite{PRL2009}, this in principle allows
to prepare a gas of $N\simeq 100$ lithium 7 atoms in a coherent superposition
of being at two different locations separated by $\simeq 100\, \mu$m.
We thus restrict to the most interesting large-$N$ limit.
As the potential barrier may be produced with a focused Gaussian laser beam, we can assume that
$U(x)$ is a repulsive Gaussian of width $b$ (the so-called waist of the laser beam):
\be
U(x) = U_0 e^{-2x^2/b^2}, \ \ \  U_0>0,
\label{eq:gauss}
\ee
so that the potential $\bar{V}(X)$ is also even and bell shaped.
We shall also assume, as in \cite{PRL2009}, that
\be
\frac{\hbar^2 \bar{K}^2}{2M} = \frac{\Delta}{2},
\label{eq:choiceK0}
\ee
so that in the large $N$ limit, 
\be
\bar{K}\sim \frac{N^{1/2}}{2\sqrt{2}\xi}.
\label{eq:choiceK02}
\ee
Having a significantly smaller $\bar{K}$ would indeed uselessly slow down the Schr\"odinger's cat formation process.
Having a significantly larger $\bar{K}$ is forbidden by the elasticity condition (\ref{eq:constraint_E}).

\noindent{\sl Case $b\gg \xi$:} This broad barrier case is experimentally the typical one,
since the waist of a focused laser beam is a few microns, whereas $\xi \lesssim 1\mu$m \cite{PRL2009}.
Then $\bar{V}(X)\simeq N U(X)$, and the width of $\bar{V}$ is also of order $b$. 
Since $\bar{K} b \propto N^{1/2} b/\xi$ is much larger than unity,
the scattering problem of the centre-of-mass on $\bar{V}(X)$ is in the semiclassical regime,
where the transmission and reflection amplitudes at incoming wavevector
$K$ have the approximate expressions, see Eqs.~(3.49,3.58,4.23) in \cite{Berry}:
\bea
\label{eq:t0sc}
t_0 &\simeq& \frac{e^{-\lambda(K)}e^{i[\alpha(K)-\beta(K)]}}{[1+e^{-2\lambda(K)}]^{1/2}} \\
r_0 &\simeq & \frac{-i\, e^{i[\alpha(K)-\beta(K)]}}{[1+e^{-2\lambda(K)}]^{1/2}} \\
\lambda(K) &=& \int_{X_-}^{X_+}  dX\ (-i) K(X) \geq 0 \\
\alpha(K) &=& 2 K X_- + 2 \int_{-\infty}^{X_-} dX [K(X)-K] \\
\beta(K) &=& \frac{\lambda(K)}{\pi}\ln \left|\frac{\lambda(K)}{\pi e}\right| \!
+\!\arg\Gamma\!\left(\!\frac{1}{2}-\frac{i\lambda(K)}{\pi}\!\right) \\
K(X) &=& \left[K^2-\frac{2M \bar{V}(X)}{\hbar^{2}}\right]^{1/2}
\label{eq:defKX}
\eea
Here $X_- \leq 0$ and $X_+=-X_-$ are the two classical turning points for an incoming energy below
the maximum $V_0$ of $\bar{V}$, and $K(X)$ is either in $\mathbb{R}^+$ or in 
$i\mathbb{R}^+$ if $X$ is in the classically allowed or forbidden region. 
For an incoming energy larger than $V_0$, one has to use analytic continuation \cite{Berry}.
The phase $\beta(K)$ is given in \cite{Berry}, and the extra phase $\alpha(K)$
is due to our different choice for the phase reference point.
We also recall the WKB approximation for $\Phi_0(X)$ in the classically allowed region:
\bea
\!\!\!\!\!\!\Phi_0(X)\!\!\!\! &\stackrel{X<X_-}{\simeq}&\!\!\!\!  [K/K(X)]^{1/2} \Big[ e^{iKX} e^{i\int_{-\infty}^X dx [K(x)-K]}
\nonumber \\
&& + r_0 e^{-iKX} e^{-i\int_{-\infty}^{X} dx [K(x)-K]} \Big]
\label{eq:wkb1} \\
\!\!\!\!\!\!\Phi_0(X)\!\!\!\! &\stackrel{X>X_+}{\simeq}& \!\!\!\! [K/K(X)]^{1/2} t_0 e^{iKX} e^{i\int_{+\infty}^{X} dx [K(x)-K]}
\label{eq:wkb2}
\eea

We then see that a transmission probability of $1/2$ is achieved in the semiclassical formula (\ref{eq:t0sc})
for an energy equal to $V_0$, that is for a momentum
\be
K_{1/2} = \frac{(2M V_0)^{1/2}}{\hbar}.
\ee
Away from this value of $K$, $|t_0|^2$ will drop rapidly to zero or rise rapidly to one.
A local formula is obtained by approximating the top of $\bar{V}(X)$ around $X=0$ by
a parabola, so that
\be
|t_0|^2 \simeq \frac{1}{1+\exp[(K_{1/2}-K)/\delta K]}
\ee
with
\be
\delta K = \frac{1}{2\pi} \left(\frac{|\bar{V}''(0)|}{2\bar{V}(0)}\right)^{1/2}.
\ee
For large $N$,  one finds $\delta K \simeq 1/(\sqrt{2}\pi b)$ so that one is experimentally
in the regime $\Delta K\gg \delta K$.  In what follows, we thus adjust the barrier height
to have $K_{1/2}=\bar{K}$. Then, with Eq.~(\ref{eq:choiceK0}),
\be
N U_0  \simeq V_0 = \frac{\Delta}{2}.
\label{eq:choiceU0}
\ee

Finally, we evaluate the parameter $\epsilon$ of (\ref{eq:epsilon}) appearing
in the bracketing (\ref{eq:bound_t},\ref{eq:bound_r}) (that can be used here since $U(x)$ and
$\bar{V}(X)$ are even), for the physically most relevant case $K=\bar{K}=K_{1/2}$.
We use the large-$N$ estimate for a broad barrier, see Eqs.~(\ref{eq:wbblne},\ref{eq:CNg}),
and the simplest estimate (neglecting rapidly oscillating terms):
\be
|\Phi_0(X)|^2 \propto \frac{K_{1/2}}{K(X)}
\ee
with $K(X)$ defined in (\ref{eq:defKX}) and a proportionality factor equal to $1+|r_0|^2=3/2$ for $X<0$ 
and to $|t_0|^2=1/2$ for $X>0$.  Since $K(X)$ vanishes linearly in the classical turning point $X=0$, we get a logarithmic divergence
in the resulting approximation for $\langle \Phi_0| W|\Phi_0\rangle$, that we cut by introducing the quantum length
scale $a_{\rm ho}\ll b$ associated to the Schr\"odinger's equation in the inverted
parabola approximating $\bar{V}(X)$ close to its maximum
\footnote{Failure of the semiclassical approximation is customary close to classical turning points,
where one usually performs a local full quantum study by linearizing the potential,
which leads to an Airy function for the wavefunction \cite{Berry}. The unusual
feature here is that the classical turning point is located at a potential maximum,
where $\bar{V}(X)$ has to be approximated by a parabola and $\Phi_0(X)$ may be expressed
locally in terms of $J_{1/4}$ and $N_{1/4}$ Bessel functions. Using these Bessel functions
for the local study of the scattering problem around $X=0$, on an arbitrary interval $X\in (-l,l)$ with
$a_{\rm ho}\ll l \ll b$, we have checked that our simple cutting procedure at a distance $a_{\rm ho}$ is correct within
logarithmic accuracy. In particular it is found that the oscillating terms for $X<0$ give rise
to an integral of the form $\int^{l/a_{\rm ho}} \frac{dx}{x} \sin(x^2)$ that converges
for $l/a_{\rm ho}\to +\infty$  and thus does not affect the logarithmically divergent bit.}:
\be
a_{\rm ho} = \frac{\hbar^{1/2}}{(M |\bar{V}''(0)|)^{1/4}} \simeq \left(\frac{b}{\sqrt{2}\bar{K}}\right)^{1/2} \simeq 
\left(\frac{2 b\xi}{N^{1/2}}\right)^{1/2}
\ee
Keeping only the logarithmically diverging contribution amounts to approximating the matrix element as
\be
\label{eq:cutting}
\langle \Phi_0| W(\hat{X})|\Phi_0\rangle \simeq 2W(0) \int_{a_{\rm ho}}^{b} dX\, \frac{b}{X\sqrt{2}},
\ee
which leads to the estimate
\be
\epsilon \simeq \left(\frac{2\xi}{b}\right)^3 \left(\frac{2}{N}\right)^{1/2} \left[\frac{\pi^2}{3}+2\zeta(3)\right] \ln (b/a_{\rm ho}).
\ee
In conclusion, for $|t_0|^2\simeq 1/2$ in the broad barrier case, we find in the large $N$ limit (where $\epsilon\ll 1$):
\be
|t-t_0| = O\left[\frac{(\xi/b)^3}{N^{1/2}}\ln\left(N^{1/2}b/\xi\right)\right],
\ee
as already given in \cite{PRL2009}.

\noindent {\sl Case $b\ll \xi$:} For a narrow barrier, 
\be
\bar{V}(X) \simeq \rho(X|0) \int_{-\infty}^{+\infty} dx\ U(x).
\ee
In the large $N$ limit, replacing $\rho(X|0)$ by its classical field approximation,
that is the leading term in the right-hand side of (\ref{eq:devVbar}), gives
\be
\bar{V}(X) \simeq \frac{V_0}{\cosh^2(X/b_{\rm eff})}
\label{eq:vbpud}
\ee
with
\bea
V_0 &=& \frac{N}{4\xi} \int_{-\infty}^{+\infty} dx\ U(x) \\
b_{\rm eff}   &= & 2\xi.
\eea
Although the resulting scattering problem for $\Phi_0(X)$ then becomes exactly solvable \cite{Landau},
we simply reuse the semiclassical reasoning of the previous (broad-barrier) case, since
$\bar{K} b_{\rm eff} \simeq  (N/2)^{1/2} \gg 1$ is again in the semiclassical regime.
At half transmission probability, 
\be
\frac{N U_0}{\Delta} \simeq  \left(\frac{8}{\pi}\right)^{1/2} \frac{\xi}{b} \gg 1 
\ee
so that $N U_0$ is now much larger than the gap $\Delta$, contrarily
to the broad barrier case.  The harmonic oscillator length used in the cutting procedure 
is found to be $a_{\rm ho}\simeq b_{\rm eff} (2/N)^{1/4}$. We use the equivalent of Eq.~(\ref{eq:cutting}) and
we estimate $W(0)$ from Eq.~(\ref{eq:w_si_pot_etroit}).
At half transmission probability,
we finally obtain from the simple bracketing (\ref{eq:bound_t},\ref{eq:bound_r}):
\be
|t-t_0| \lesssim  \frac{\xi/b}{N^{1/2}}  \frac{\ln(N/2)}{(2\pi)^{1/2}}.
\ee
As expected, this bound diverges for $b\to 0$ (at fixed $N$), since $U(x)$ then approaches a Dirac
potential, for which the use of the improved bracketing (\ref{eq:boundimp_t},\ref{eq:boundimp_r})
is more appropriate and leads at half transmission for large $N$ to
\be
|t-t_0| \lesssim \frac{\ln(N/2)}{8 N^{1/2}},
\label{eq:bscd}
\ee
where Eq.~(\ref{eq:Wimpdelta}) was used with $v=U_0 b\, (\pi/2)^{1/2}$.

\noindent{\sl The Born-Oppenheimer prediction:}
For the delta external potential $U(x)=v\delta(x)$, it is interesting to compare the
upper bound (\ref{eq:bscd}) to the result of Sec.~\ref{sec:BO}. In the present large
$N$ limit, one can treat $\mathcal{V}_{\rm BO}$ with first order perturbation 
theory similarly to Eqs.~(\ref{eq:dtpo},\ref{eq:drpo}) and one can use
the expression (\ref{eq:vbopud}) for $\mathcal{V}_{\rm BO}$.
In the resulting perturbative expression for $t-t_0$, one can approximate the various quantities
by their $N\to +\infty$ limit, in particular the scattering wavefunction $\Phi_0(X)$ 
at incoming wavevector $K$ may be replaced by 
the scattering wavefunction $\Phi_0^{(0)}(X)$ of the $1/\cosh^2$ potential of Eq.~(\ref{eq:vbpud}),
which is exactly expressed in terms of an hypergeometric function \cite{Landau},
with the transmission amplitude
\be
t_0^{(0)} = \frac{\Gamma(\frac{1}{2}+is-iKb_{\rm eff})\Gamma(\frac{1}{2}-is-iKb_{\rm eff})}
{\Gamma(1-iKb_{\rm eff}) \Gamma(-iKb_{\rm eff})}
\label{eq:t00}
\ee
with
\be
s = \left(\frac{2M b_{\rm eff}^2}{\hbar^2} V_0 -\frac{1}{4}\right)^{1/2}.
\ee
Here, to zeroth order in $1/N$,
we have $K b_{\rm eff}\simeq (N/2)^{1/2}$ as in Eq.~(\ref{eq:choiceK02}),
and $s\simeq K b_{\rm eff}$ due to the half-transmission condition 
$|t_0^{(0)}|\simeq |t_0|=1/\sqrt{2}$,
so that $N m v^2/\hbar^2 \simeq \hbar^2/(16 M \xi^2)$. Further expressing
Eq.~(\ref{eq:vbopud}) in terms of the variable $\theta=\tanh(X/b_{\rm eff})$,
we obtain
\be
t-t_0\simeq \frac{i}{8(N/2)^{1/2}} \int_{-1}^{1} d\theta 
(1-\frac{3}{2}\theta^2+\frac{3}{2}\theta^4) \Phi_0^{(0)}(X) \Phi_0^{(0)}(-X).
\label{eq:tmtopbo}
\ee
The terms proportional to $\theta^2$ and $\theta^4$ vanish in $\theta=0$, which allows
to directly use the WKB forms (\ref{eq:wkb1},\ref{eq:wkb2}): at half-transmission,
$K(X)=K|\theta|$ so that
\be
\Phi_0^{(0)}(X) \Phi_0^{(0)}(-X) \simeq \frac{t_0^{(0)}}{\theta}
\left(1-\frac{i}{\sqrt{2}} e^{-iKb_{\rm eff}\ln(1-\theta^2)}\right),
\label{eq:phphiwkb}
\ee
and one finds as expected that the rapidly oscillation bit gives a negligible 
contribution to the integral.
Note that the semiclassical approximation gives 
\be
t_0^{(0)}\simeq \frac{e^{-2iK b_{\rm eff}\ln 2}}{\sqrt{2}}
\ee
in agreement with the Stirling asymptotic equivalent of (\ref{eq:t00}) for
$s=K b_{\rm eff}\to +\infty$.

On the contrary, the constant term in between the parenthesis in Eq.~(\ref{eq:tmtopbo})
cannot be treated  with the simple WKB approximation (\ref{eq:phphiwkb}): As already discussed
above, this simple approximation is inaccurate over the interval $|X|\lesssim a_{\rm ho}$,
where it would incorrectly lead to a logarithmic divergent integral. As a straightforward
alternative to more elaborate semiclassical methods, we can use the fact here that an
infinitesimal change $\delta V_0$ of the amplitude $V_0$ of the $1/\cosh^2$ potential
will lead to change of $t_0^{(0)}$ that may either be evaluated by perturbation
theory, or by taking the derivative of Eq.~(\ref{eq:t00}) with respect to $V_0$, that is
with respect to $s$. This leads to the exact relation
\begin{multline}
\int_{-1}^{1} d\theta \Phi_0^{(0)}(X) \Phi_0^{(0)}(-X)  = -\frac{Kb_{\rm eff}}{s}
t_0^{(0)} [\psi(\frac{1}{2}+is-iKb_{\rm eff}) \\
-\psi(\frac{1}{2}-is-iKb_{\rm eff})]
\end{multline}
where the digamma function $\psi(z)$ behaves as $\ln z+o(1)$ for $|z|\to +\infty$.

In conclusion, for the soliton scattering
at a centre-of-mass kinetic energy $\Delta/2$ on a delta external potential
such $|t_0|=1/\sqrt{2}$,  the Born-Oppenheimer-like approach predicts that, in the large
$N$ limit with $N g$ fixed,
\be
\frac{t-t_0}{t_0} \sim  \frac{i}{8(N/2)^{1/2}} \left[\frac{1}{2}\ln(2N) -\psi(1/2) 
-i\frac{\pi}{2} -\frac{3}{4}\right],
\label{eq:exactpde}
\ee
where we recall that $\psi(1/2)=-2\ln 2 -C$ and $C=0.57721\ldots$
is Euler's constant.
This is compatible with the bound (\ref{eq:bscd}).
Since the equivalence conditions of the Born-Oppenheimer-like approach with the 
systematic Bogoliubov approach of Sec.~\ref{sec:lnl} are here satisfied, as discussed
in the paragraph below Eq.~(\ref{eq:vbopud}), the typical centre-of-mass wavevector
diverging as $N^{1/2}$, the result (\ref{eq:exactpde}) is 
asymptotically exact
\footnote{One can also treat the deviation of $\bar{V}(X)$
from Eq.~(\ref{eq:vbpud}) to first order in pertubation theory to
obtain an asymptotic equivalent of $t_0-t_0^{(0)}$. This, combined with
Eq.~(\ref{eq:exactpde}), leads to
$(t-t_0^{(0)})/t_0\sim \frac{9i}{32(N/2)^{1/2}}$, without any $\ln N$ contribution,
due to the choice (\ref{eq:choiceK0}).}.

\subsection{Application of the improved bracketing to $N=2$}

Explicit calculations of the improved error potential $W_{\rm imp}(X)$ of Eq.~(\ref{eq:Wimp})
may be performed for $N=2$, that is for the scattering of a dimer, on a delta barrier $U(x)=v\delta(x)$, $v>0$.
In this case, the set of internal coordinates reduce to the relative coordinate $x=x_2-x_1$ of the two particles,
with $x_1 = X -x/2$ and $x_2=X+x/2$. The internal Hamiltonian is simply
\be
H_{\rm int} = -\frac{\hbar^2}{m} \frac{d^2}{dx^2} + g \delta(x).
\ee
Its normalized ground state wavefunction is $\phi(x)=q_0^{1/2}  e^{-q_0 |x|}$, with an energy $E_0(2)= -\hbar^2 q_0^2/m$ in
agreement with Eq.~(\ref{eq:E0N}),
where we have set 
\be
q_0 = -\frac{m g}{2\hbar^2}.
\ee
This immediately gives the mean potential
\be
\bar{V}(X) = 4 v |\phi(2X)|^2 = 4 v q_0 e^{-4 q_0 |X|}.
\ee
Since the continuous spectrum of $H_{\rm int}$ starts at zero energy, one has the gap $\Delta=-E_0(2)$, in agreement
with Eq.~(\ref{eq:DeltaN}).
The Green's function of $H_{\rm int}$ at energy 
\be
E=-\frac{\hbar^2 q^2}{m}, \ \ \ 0<q<q_0,
\ee
is also easily calculated from
the differential equation
\be
[E+\frac{\hbar^2}{m} \partial_x^2 - g \delta(x)] \langle x |(E-H_{\rm int})^{-1} |y\rangle = \delta(x-y)
\ee
with the boundary conditions that it does not diverge exponentially for $|x|\to +\infty$. E.g.\ for $y <0$, one simply has to integrate
the differential equation over $x$ over the intervals $x<y$, $y < x < 0$, $0 < x$, where the general solution is the sum
of two exponential functions of $x$, and then match the solutions in $x=y$ and $x=0$, using the continuity of the Green's function
with $x$, and the discontinuity
of its first order derivative with respect to $x$ as imposed by the Dirac terms. To finally obtain the matrix elements of
$\mathcal{Q}/(E\mathcal{Q} -\mathcal{Q} H_{\rm int} \mathcal{Q})$ in position space, one simply has to remove from the Green's function of $H_{\rm int}$
the contribution $\phi(x) \phi(y)/[E-E_0(2)]$ of the ground state of $H_{\rm int}$.
We finally obtain
\begin{multline}
W_{\rm imp}(X) = \frac{4 m v^2q_0}{\hbar^2 q} e^{-4 q_0 |X|} \left[1 - \frac{q_0+q}{q_0-q} e^{-4 q |X|}  \right .\\
\left.+ \frac{4 q q_0}{q_0^2-q^2} e^{-4 q_0 |X|} \right].
\end{multline}
A numerical or an analytical solution \footnote{An analytical solution can be obtained in
terms of Bessel functions after an exponential change of variable.}  
of the scattering problem for $\Phi_0(X)$ can then be combined to this expression
for $W_{\rm imp}(X)$, to obtain explicit numbers for the improved bracketing
(\ref{eq:boundimp_t},\ref{eq:boundimp_r}).

\section{Conclusion}
\label{sec:conclusion}

We have considered the scattering of a one-dimensional quantum soliton 
(the bound state of $N$ attractive-$\delta$ bosons) on a potential barrier
in the {\sl elastic} regime, where energy conservation prevents
observation of soliton fragments at infinity. The scattering at a given
incoming centre-of-mass wavevector $K$ is then characterized by 
reflection and transmission amplitudes $r$ and $t$ for the
soliton centre-of-mass wavefunction $\Phi(X)$, with $|t|^2+|r|^2=1$.

In the simplest approximation, one assumes that $\Phi(X)$
simply sees a local potential $\bar{V}(X)$ obtained by averaging the
single particle external potential $U(x)$ over the particle density
profile of the quantum soliton, leading to approximations $r_0$
and $t_0$ for the amplitudes $r$ and $t$. 
Rigorous upper bounds on the resulting errors
$|r-r_0|$ and $|t-t_0|$ are derived and are expressed
in an operational form (distinguishing various limits of broad
or narrow barrier, for any $N$ or for large $N$).

In an exact treatment, also giving a precise meaning to $\Phi(X)$,
it is shown that an additional, non-local potential $\delta \mathcal{V}$
appears in an effective Schr\"odinger's equation for $\Phi$. 
The large $N$ leading behaviour for $\delta\mathcal{V}$ is obtained using
Bogoliubov theory, and it is compared to a Born-Oppenheimer-like approach
that treats the centre of mass of the system as the heavy particle.

Finally, simple applications of the formalism are given, mainly in
the context of the Schr\"odinger cat state production scheme considered in
\cite{PRL2009,Cederbaum}.

\begin{acknowledgments}
C.W. thanks the UK EPSRC for funding (Grant No.  EP/G056781/1) and the EU for financial support
during his stay in Paris (contract MEIF-CT-2006-038407).
The group of Y.C. is a member of IFRAF and acknowledges financial support from IFRAF.
We acknowledge a discussion with A. Sinatra.
\end{acknowledgments}

\appendix
\section{Bounds on the error on $t$ and $r$}
\label{appen:theorem}

In this Appendix we shall prove Eqs.(\ref{eq:bound_tg},\ref{eq:bound_rg}).
We rewrite Eq.~(\ref{eq:heff}) in position space,
\be
\frac{\hbar^2 K^2}{2M} \Phi(X) = -\frac{\hbar^2}{2M} \Phi''(X)
+\bar{V}(X) \Phi(X) +S(X)
\label{eq:with_source}
\ee
with the contribution that we shall treat formally as a source term,
\be
S(X) = \langle X| \delta \mathcal{V}|\Phi\rangle.
\ee
Here $\Phi(X)$ obeys the boundary conditions (\ref{eq:alinfinip},\ref{eq:alinfinim}),
and the fact that $\delta\mathcal{V}$ is hermitian leads to $|r|^2+|t|^2=1$ as expected
\footnote{One multiplies Eq.~(\ref{eq:with_source}) by $\Phi^*(X)$ and the complex conjugate
of Eq.~(\ref{eq:with_source}) by $\Phi(X)$, and one makes the difference between the two resulting
equations, that one integrates over $X$ from $-\infty$ to $+\infty$, using
$\Phi^*(X)\Phi''(X)-\Phi(X)\Phi''^{*}(X)=\frac{d}{dX}[\Phi^*(X)\Phi'(X)-\Phi(X)\Phi'^{*}(X)]$.}.
To integrate formally Eq.~(\ref{eq:with_source}) we need two independent
solutions of the corresponding homogeneous equation. One is
$\Phi_0(X)$, i.e.\ the scattering solution for an incoming
wave from the left (i.e.\ with a centre-of-mass wavevector $K>0$). 
The other solution is conveniently taken
as the scattering solution for a wave incoming from the right
(i.e.\ with a centre-of-mass wavevector $(-K)<0$), denoted as
$\tilde{\Phi}_0(X)$.
If the potential $U(x)$ is even, we simply have
$\tilde{\Phi}_0(X)=\Phi_0(-X)$. In the general case,
we take Eq.~(\ref{eq:phi0t}).
Then one may check that
\bea
\tilde{\Phi}_0(X) &\sim& 
e^{-iKX} -\frac{r_0^* t_0}{t_0^*} e^{iKX} \ \ \mbox{for}\ \ X\to +\infty \\
\tilde{\Phi}_0(X) &\sim& 
t_0 e^{-iKX}\ \ \mbox{for}\ \ X\to -\infty.
\eea
Then, after formal integration with the method of variation of constants
and calculation of the Wronskian of $\Phi_0(X)$
and $\tilde{\Phi}_0(X)$,
\be
\mathcal{W}(X) = \Phi_0(X) \tilde{\Phi}_0'(X)-\Phi_0'(X) \tilde{\Phi}_0(X) = -2iKt_0,
\ee
we obtain
\be
\Phi(X) = [1+A(X)] \Phi_0(X) + B(X) \tilde{\Phi}_0(X)
\label{eq:sol_for}
\ee
with
\bea
A(X) &=& -\frac{iM}{\hbar^2 K t_0} \int_{-\infty}^X dx\, \tilde{\Phi}_0(x) S(x) \\
B(X) &=& \frac{iM}{\hbar^2 K t_0} \int_{+\infty}^X dx\, \Phi_0(x) S(x).
\eea
One may check that $\Phi(X)$ obeys the right boundary conditions
(\ref{eq:alinfinip},\ref{eq:alinfinim}) with
\bea
\label{eq:t_t0}
t &=& t_0 [1+A(+\infty)] \\
r &=& r_0 + t_0 B(-\infty).
\label{eq:r_r0}
\eea
An upper bound of $|A(X)|$ is obtained from Eq.~(\ref{eq:maj}) by taking
\bea
\Phi_1(x) &=& \theta(X-x) \tilde{\Phi}_0^*(x)  \\
\Phi_2(x) &=& \Phi(x),
\eea
where $\theta$ is the Heaviside step function.
Furthermore we use the fact that
\be
\langle\Phi_1| W(\hat{X})|\Phi_1\rangle \leq \langle \tilde{\Phi}_0 | W(\hat{X})| 
\tilde{\Phi}_0\rangle,
\ee
since $W(X)$ is positive.
Then
\be
|A(X)| \leq \epsilon \eta^{1/2} \alpha^{1/2},
\label{eq:majA}
\ee
where $\epsilon$ is defined in Eq.~(\ref{eq:epsilon}),  $\eta$ is defined
in Eq.~(\ref{eq:eta})
and
\be
\alpha =\frac{\langle \Phi| W(\hat{X})|\Phi\rangle}
{\langle \Phi_0| W(\hat{X})|\Phi_0\rangle}.
\ee
Similarly, taking $\Phi_1(x) = \theta(x-X) \Phi_0^*(x)$, one has
\be
|B(X)| \leq \epsilon \alpha^{1/2}.
\label{eq:majB}
\ee
The last step is to derive an upper bound for $\alpha$. To this end, 
we derive an upper bound on $|\Phi(X)|$ from Eq.~(\ref{eq:sol_for}):
\be
|\Phi(X)| \leq |\Phi_0(X)| + \epsilon \alpha^{1/2} [\eta^{1/2}|\Phi_0(X)|+
|\tilde{\Phi}_0(X)|].
\ee
We square this relation, multiply by the positive quantity $W(X)$,
integrate over $X$, divide by $\langle \Phi_0| W(\hat{X})|\Phi_0\rangle$
and use the Cauchy-Schwarz inequality 
\begin{multline}
\int_{-\infty}^{+\infty} dX\, |\Phi_0(X)| |\tilde{\Phi}_0(X)| W(X) 
\leq \eta^{1/2}\\
\times \int_{-\infty}^{+\infty} dX\, |\Phi_0(X)|^2 W(X)
\end{multline}
to obtain
\be
\alpha \leq 1 + 4\epsilon\eta^{1/2} \alpha^{1/2} + 4\epsilon^2\eta\alpha = 
\left(1+2\epsilon \eta^{1/2}\alpha^{1/2}\right)^2.
\ee
Taking the square root leads to
\be
\alpha^{1/2} \leq 1 + 2 \epsilon\eta^{1/2} \alpha^{1/2}.
\ee
For $\epsilon\eta^{1/2} < 1/2$ we thus get
\be
\alpha^{1/2} \leq \frac{1}{1-2\epsilon\eta^{1/2}}.
\ee
This inequality, together with 
Eqs.(\ref{eq:t_t0},\ref{eq:r_r0},\ref{eq:majA},\ref{eq:majB}), 
leads to Eqs.(\ref{eq:bound_tg},\ref{eq:bound_rg}).
In the particular case of an even external potential $U(x)$, $W(X)$ is
also even and one has $\eta=1$, which leads to the simpler relations
Eqs.(\ref{eq:bound_t},\ref{eq:bound_r}).


\begin{thebibliography}{99}
\bibitem{help}
This is the usual notation,
implying that the inversion of an operator $\mathcal{Q} A\mathcal{Q}$ has to be understood
as the inversion, within the subspace over which $\mathcal{Q}$ projects,
of the restriction of $A$ to that subspace.
\end{thebibliography}
\end{document}